 \documentclass[preprint, english,eqsecnum,aps,pre,showpacs]{revtex4}

\usepackage[T1]{fontenc}
\usepackage[latin1]{inputenc}
\usepackage{amsmath}
\usepackage{amssymb}

\makeatletter
\usepackage{graphicx}   
\usepackage{dcolumn}    
\usepackage{bm}         
\usepackage{babel}
\begin{document}

\title{Relaxation times under pulsed ponderomotive forces and the Central Limit Theorem}

\author{J.L.~Domenech-Garret $^{1}$}
\email{domenech.garret@upm.es}

\affiliation{$^{1}$ Departamento de F\'{\i}sica. 
ETSI Aeron\'autica y del Espacio.\\
\\ 
Universidad Polit\'{e}cnica de Madrid, 28040 Madrid, Spain.}

\date{\today}

\begin{abstract}
We study the relaxation time of a generic plasma which is perturbed by means of a  time-dependent pulsed force. This time pulse is modelled using a Gaussian superposition. During such a pulse two forces are considered: An inhomogeneous oscillating electric force and the corresponding ponderomotive force. The evolution of that ensemble is driven by the Boltzmann Equation, and the perturbed population is described by a power-law distribution function. In this work, as a new feature, instead the usual techniques the transient between both distributions is analysed using  the moments of such distribution function and the Central Limit Theorem. This technique, together with the, ad hoc solved, equation of motion of the charges under this particular system of pulsed forces, allows to find the corresponding expressions relating the time pulse with the  relaxation times and the dynamic  conditions. We validate that new technique by comparison with the  analytical expression using the corresponding relaxation time using an exact collision  operator. Moreover, we parameterise this plasma to make  numerical estimates in order to  analyse  the impact of relevant  parameters involved in the physical process on such a relaxation time.   
\end{abstract}


\maketitle

\section{\textbf{Introduction}}
\label{intro}
The Boltzmann equation (BEq hereafter)  is  used to analyse many problems on transport and kinetics of a collection of particles in non-equilibrium statistical mechanics. One of the paradigmatic fields of application is the physics of  partial and fully ionised plasmas.  There are a lot of excellent works dealing on Beq in the literature (see, for instance \cite{Chapman, Raizer,Lieberman,Chen},  among many others) and its applications: For example, the effect of electron collisions on the propagation of radio waves in the ionosphere, shock waves, and many other problems. The equilibrium Maxwell, $f_0$ distribution (or the Fermi-Dirac in quantum problems) are exact solutions of the Boltzmann collision integral within the Beq. For states in  which the population deviate slightly from equilibrium,  they are represented usually as a distribution function $(f_0 + f_1+ \ldots)$, where $f_1$ represents a correcting term, \cite{Zhdanov}. Another approach on the fundamental problems on statistical mechanics is to use the Uhlenbeck-like collision term derived a consequence of the Fokker-Planck equation \cite{Alekseev}. In plasma Physics, one of the  most widely used collisional operators to include into the BEq  is the so called the Bhatnagar-Gross-Krook (BGK) collision operator, arising from the general case  by using a mean-free-path treatment, \cite{BGK}. It represents, on the average, that the electron is only free for $\tau_c$ seconds before it is scattered. We will discuss in detail about this operator in further sections.

As mentioned, there are many ways in the literature to represent a non-equilibrium distribution, $f$. Leaving aside the above mentioned approaches, a explicit form of the non-equilibrium distribution widely used is the so called power-law distribution. For instance, we can find it in  classical collisions, non-Maxwellian distribution functions for ions of a radio frequency ion trap, \cite{Zhakarov, Kats1, DeVoe}, or aggregation systems with injection \cite{Takayasu}. Also, the power-law, among many fields, is considered within astrophysical problems as solar flares \cite{SolarFlares}.\\

This work studies the relaxation time of a plasma which is perturbed by means of a  time-dependent pulsed force. This time pulse is modelled using a Gaussian superposition. During such a pulse two forces acting on the charges  are considered, one arising from an intense  inhomogeneous  oscillating electric field, the other one corresponds to the  ponderomotive force. The ponderomotive force acts on the plasma particles and it causes the charges to move towards the region of the weak field strength \cite{Morales}. The resulting force acting on the  plasma, if the field is able to grow enough, it may alter the initial plasma parameters, the so called regime of strong turbulence. The theory is an active field  under development, and its importance is unclear under astronomical circumstances. \cite {Chen, Benz6}. In addition, the relaxation time both of the ion and  electron beam distributions is an open problem in astrophysics, as solar radio bursts   \cite{Benz6, Zharkova}. Moreover, there are a lot of problems in space physics  involving several kind of ponderomotive forces, \cite{Lundin}: The so called Abraham force, which is proportional to the time derivative of the square of the electric wave amplitude. The  Barlow force, proportional to the product of the collision frequency with the square of the electric wave amplitude. The  Magnetic Moment Pumping, whis arises from an inhomogeneous magnetic field.  and Finally the Miller force, widely used in space physics, which  is the ponderomotive force we use in this effort. The so called Miller force, the ponderomotive force  hereafter, as we see later, comes from an inhomogeneous oscillating electric field $\textbf{$\mathcal{E}$}(x,t)$  with frequency $\omega$ and it is based on the microscopic appproach by averaging  over time the Lorentz force (see, for instance, \cite{Chen}). This  force in terms of the plasma frequency $\omega_p$ can be written as: $F_{Pdm}= -(\omega_p^2/2\omega^2) \nabla(\epsilon_0 \textbf{$\mathcal{E}$}^2)$, being $\epsilon_0$ the dielectric constant. A pedagogical  and comprehensive study on these ponderomotive forces can be found in \cite{Lundin}.  Besides, the ponderomotive force is itself  an active field of study \cite{Dodin1, Dodin2, Kazhanov}. Moreover, the ponderomotive force is present in applications like: Tokamaks, particle accelerators, ion traps, plasma thrusters \cite{Bathgate}, and laser fusion\cite{Tajima}. A study on the ponderomotive forces in laser and its applications can be found in \cite{Hora}.\\ 
In this study, the physical scenario could be a space plasma perturbed by an intense, oscillating, inhomogeneous electric field. This circumstance has been extensively studied in astronomical plasmas  \cite{Lundin}. The study of the evolution of the charged population is performed using the Boltzmann equation, where, as a novelty, instead of the usual techniques, we will model the collision term using the Central Limit Theorem. The pulsed force involves the ponderomotive force,  and here  is considered in a general plasma. The numerical estimates we will perform here only seek to study the impact of relevant physical parameters on the obtained expression for the relaxation times. A simulation describing the complete physics of an actual spatial plasma is beyond the scope of this work. This would entail defining a complete set of parameters for that specific plasma, the actual boundary conditions, and, in addition, the actual conditions of the external pulsed electromagnetic field that perturbs that plasma. As we will use a power-law distribution function for the perturbed population, concerning the collision integral, here we consider the plasma population  with a source and a sink to ensure ensure the energy flow along the spectrum in momentum space\cite{Zhakarov}. 
The choice made here of a power-law distribution function representing the non-equilibrium population is made within the framework of the usual Boltzmann-Gibbs (BG) statistics. However, there are also studies that describe these non-equilibrium states in more general contexts, such as the so-called q-Gaussian, within the framework of non-extensive statistics \cite{Tsallis1}. This type of distributions is formally handled by the so-called q-algebra, \cite{Tsallis2}, which depends on a parameter, $q$, and the usual BG statistics is recovered when $q=1$. As will be discussed later, this type of distributions could be of interest by seeking an extension of the present work, considering long-range correlations. Next, since the power-law distribution function representing the non-equilibrium states is taken here, some properties of such a distribution are reviewed below.

\section{The power-law distribution}
 The power-law  distribution, in terms of the particle momentum,  reads: $f(p,\mu) = C (p/p_0)^{-(2 \mu)}$. Such a distribution, from the viewpoint of Statistics is right-skewed, taking into account long tails. Here, $C$ stands for the normalisation constant and, as we will later discuss,  $p_0$ is the minimum $p$ value. to make calculations easier, hereafter we will set \mbox{$z\equiv p/p_0$}; as the dimensionless momentum. The minimum value of $z$ is $z_0 = 1$. This change makes  $f(p,\mu)$ into $f(z,\mu)$, which now reads:
\begin{equation}
f(z,\mu) = C \ z^{-2 \mu} 
\label{eq:PLDist}
\end{equation}
\noindent  
 Hereafter, we set the norm to the number of the degrees of freedom, (the particle density), $N_f$. Hence, using the Jacobian, the standard calculation of such a norm, for $f(z,\mu)$, the value \mbox{$C = (2 \mu - 1) N_f$} is easily obtained, with $(2\mu > 1)$ in order to be well defined. As it will  also be useful in this work, we need to find the $k-th$ moment of the distribution, $M_k(z)$, defined as,
    
\begin{eqnarray}
M_k(z)= \int^{\infty}_{1}  z^k f(z,\mu)\ dz = 
\frac{2 \mu - 1}{2 \mu -k -1} N_f
\label{eq:PL-Mom}
\end{eqnarray}
\noindent
Where, in order to be $M_k(z)$ well defined, the constraint $k<2\mu-1$ comes up. In the next section we provide some details about the forces to be considered for the pulses acting on the charged population.   

\begin{figure}
\includegraphics[width=7.5cm]{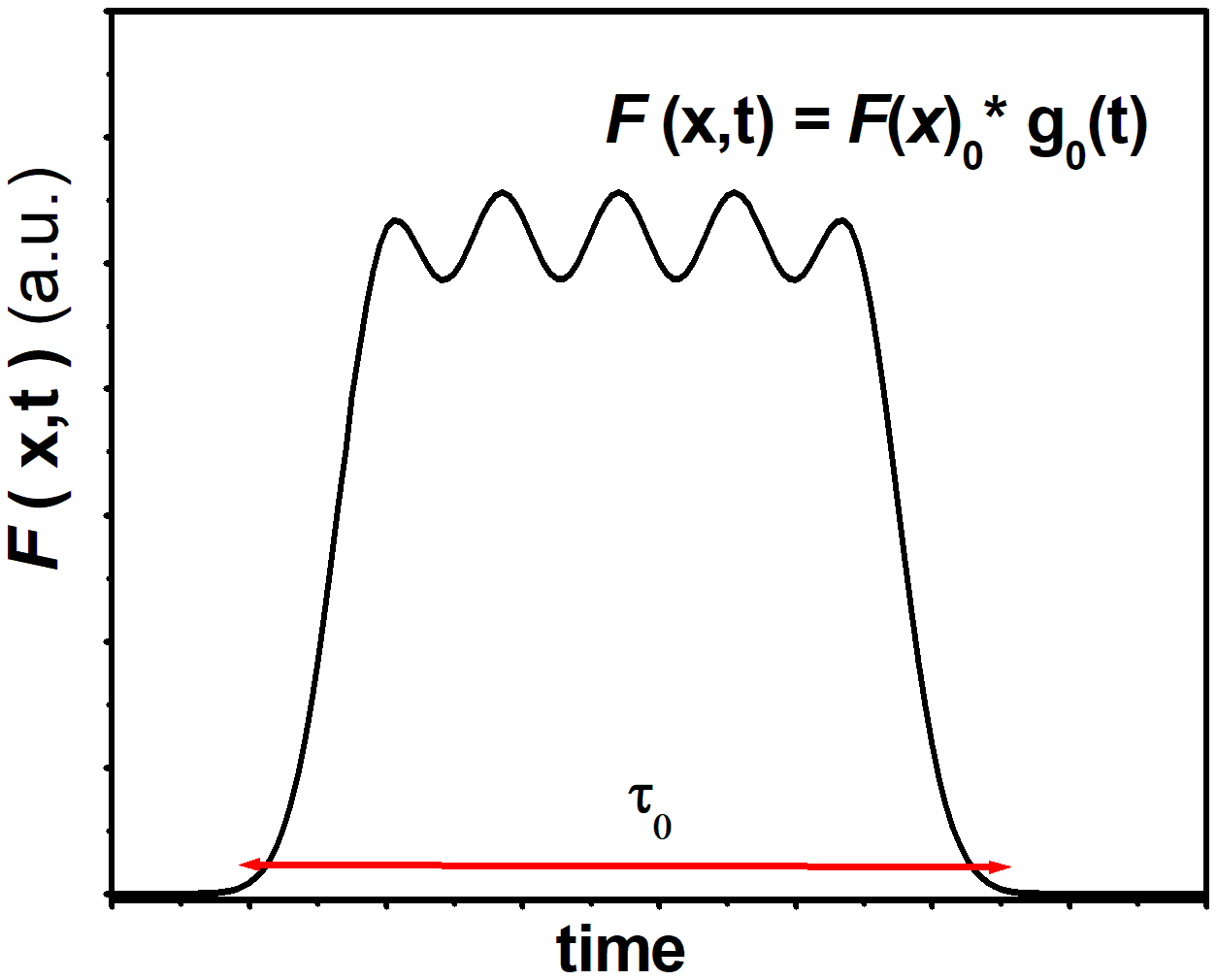} 
\includegraphics[width=7.5cm]{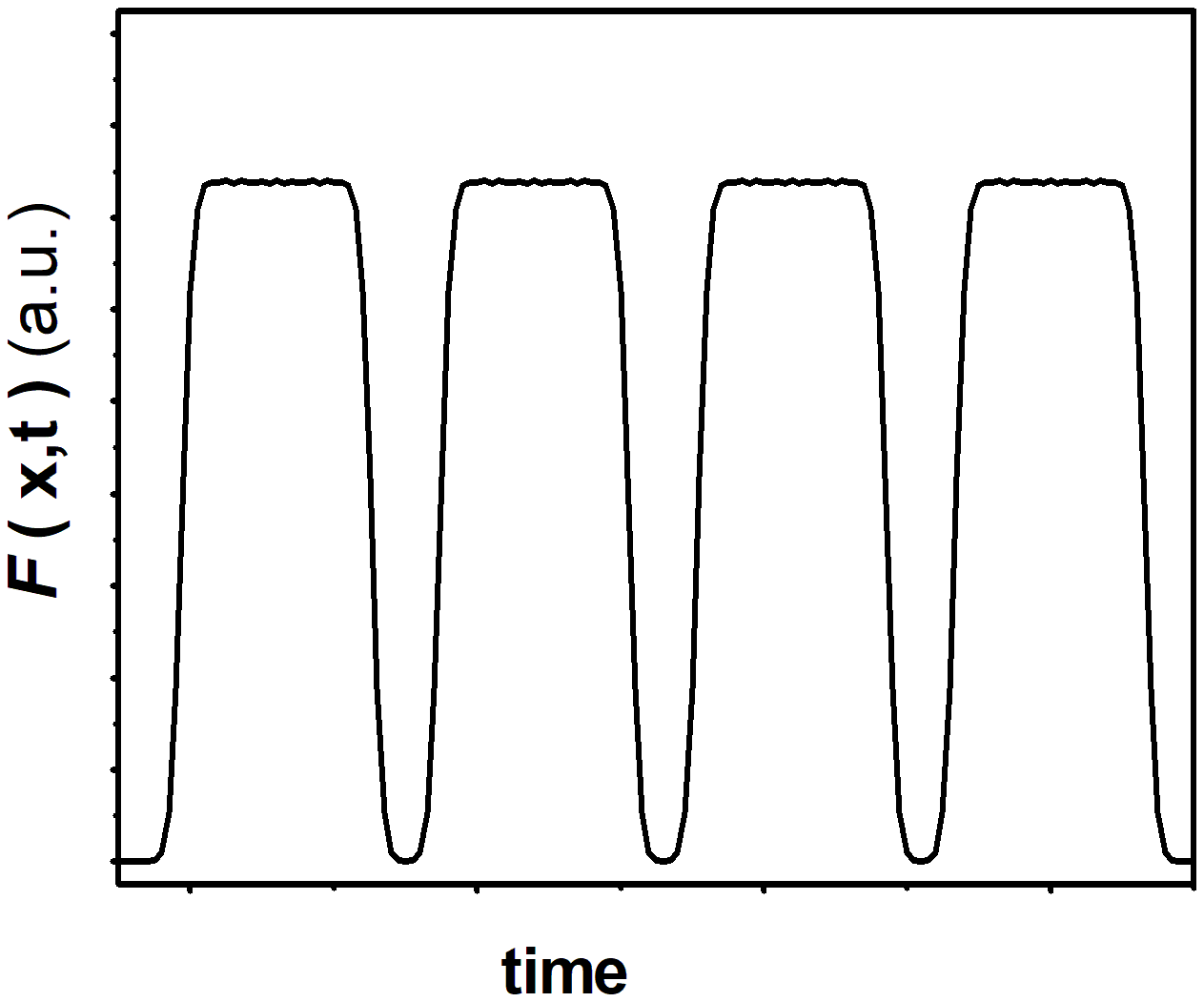} 
\caption{\label{fig1:Pulse} (A) The shape of a square pulse $\mathcal{F}(x,t)$ following $F_0(x)\  g_0(t)$ along the time $\tau_0$ ; (B) A train of square pulses modelled using the Gaussian superposition.}
\end{figure} 

\section{Square Pulse and the involved forces during the pulse}
The shape of the force, $\mathcal{F}(x,t)$, causing the departure from equilibrium to be introduced into the evolution equation can be  modelled as a square time pulse as:

\begin{eqnarray}
\label{eq:Pulse}
\mathcal{F}(x,t)\ &=& F(x)\  g_0(t); \! \! \! \! \! \\ \nonumber
g_0(t) &\equiv& \sum_{k=0}^N Exp\left[ -\left(\frac{t - 2kt_0}{t_0}\right) ^2\right]
\end{eqnarray}
\noindent
The expression of above reproduces a  local pulse of value $F_0(x)$ which holds during a given total time $\tau_0=2 N t_0$. The time factor $g_0(t)$ is modelled taking a Gaussian  superposition.  In Figure \ref{fig1:Pulse}-(A) is depicted the shape of a  square pulse  using that choice. In Figure \ref{fig1:Pulse}-(B)  a train of such a pulses in a periodical distribution can be seen. The square shape of $\mathcal{F}(x,t)$ comes from the  $g_0(t)$ factor, and as we shall see, during that time pulse, $\tau_0$, we introduce the involved forces $F(x)$ acting upon the plasma. Here, we must point out that, since  in our system actually the local force $F(x)$ will be  itself time dependent, the shape of the pulse according to Eq.(\ref{eq:Pulse}) will be modified. For simplicity, we will consider a one dimensional problem, and we set that force pointing along the \textbf{OX} axis.\\

\begin{figure}
\centering
\includegraphics[width=8.0cm]{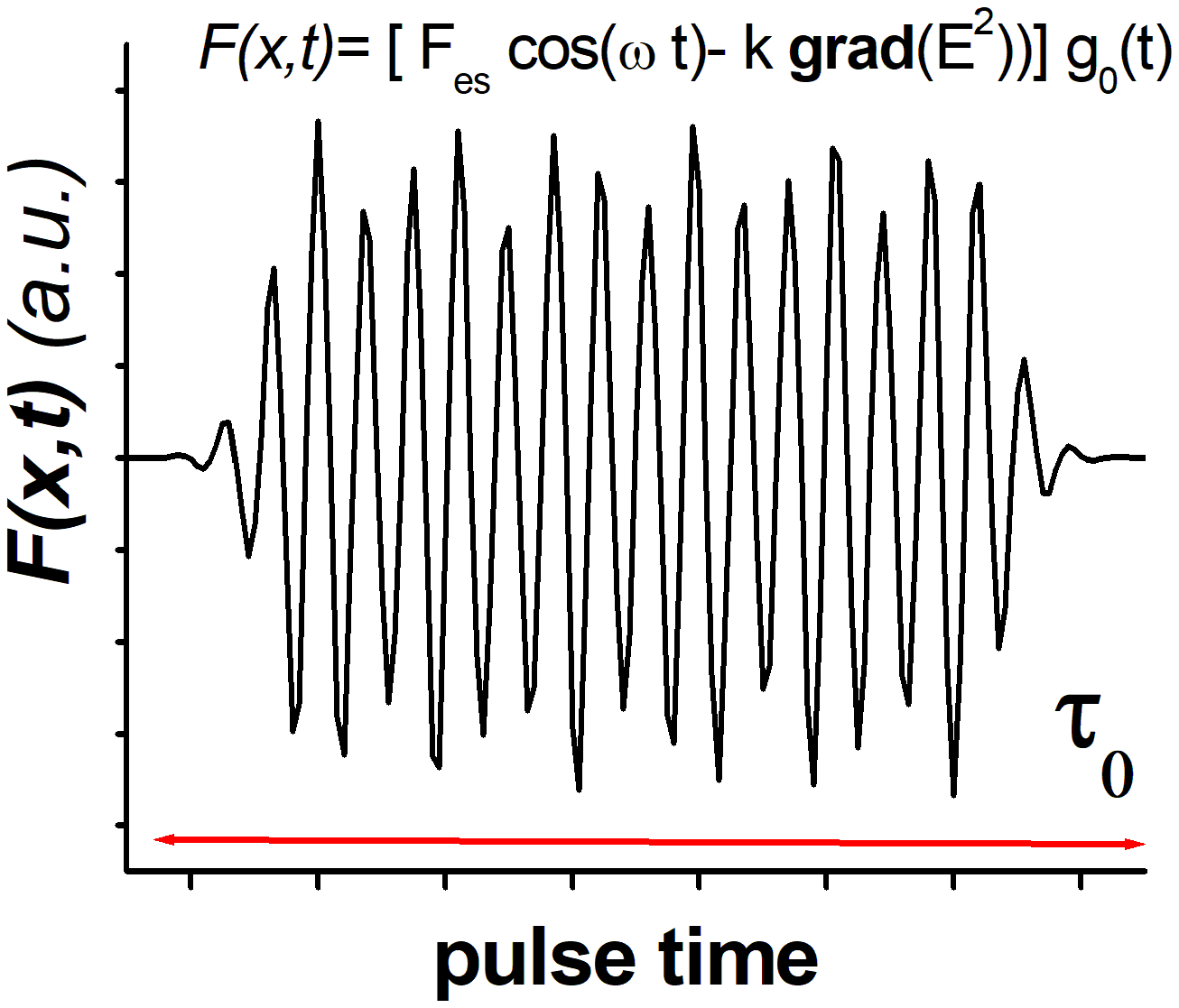}  
\caption{\label{fig2:Fe-Pond-Pulse} 
Oscillating and ponderomotive forces acting on the charged particles during the time $\tau_0$ governed by the square pulse $g_0(t)$ (in arbitrary units). Here, $k=( \omega_p^2 / 2 \omega^2 )$, and  the local $F_{es}(x)$ electric force comes from an inhomogeneous electric field.}
\end{figure} 
Next, let us analyse the involved forces acting on the charged particles of the plasma during the time $\tau_0$ governed by the square pulse above depicted. During such a time, the plasma is perturbed by means of an oscillating field, $F_{es}(x)\ cos\left(\omega t \right)$.  The local $F_{es}(x)$ electric force comes from an inhomogeneous electric field, $\mathcal{E}_{es}(x)$, along the \textbf{OX} direction. The ponderomotive force, $F_{pdm}$ hereafter, acts always in the opposite direction of the gradient of the squared electric field. The pulsed force can then be written as,    

\begin{eqnarray}
\mathcal{F}(x,t) =  \left[ F_{se}(x)\ cos\left( \omega t \right) - F_{pdm}\right]\  g_0(t); \! \! \! \! \! 
\label{eq:Forces1}	
\end{eqnarray}
In Figure \ref{fig2:Fe-Pond-Pulse} can be observed the effect of such a Gaussian pulse governing both the oscillating and the ponderomotive  force.  

\section{The Boltzmann Equation and the Central Limit Theorem}
First, let us introduce the relationship among dimensionless and the physical parameters involved hereafter:  As mentioned, we transform the power-law distribution in terms of momentum to the dimensionless momentum \mbox{$z\equiv p/p_0$}, where $p_0$ is the minimum $p$ value. We assume that such a minimum value can be taken as the thermal momentum related with the thermal energy 
\mbox{$E_{th}=1/\beta=k_B T=p_0^2/2m$}, where $k_B$ is the Boltzmann constant, and $T=T(x)$ stands for the plasma temperature, which  it is a local value. Hence, we write:

\begin{equation} 
p =\sqrt{2m/\beta}\ z  
\label{eq:pz}
\end{equation}

\section{The Boltzmann Equation} 

The BEq to be used  throughout this work can be written in a general form as: 

\begin{equation}
\label{eq:BEq-gral}
\textsl{L}\left[f(z)\right] = \textsl{I}_C\left[ z \right] 
\end{equation}
\noindent
where $\textsl{I}_C\left[ \ \ \right]$ stands for the integral collision operator. The $\textsl{L}\left[\  \right]$ operator  upon the distribution function, $f$,  corresponds to the usual derivative of the  BEq, $df/dt$ \cite{Alekseev,Kats1,Klimontovich,Lindsay}. The expansion of that derivative includes the acting force, $\mathcal{F}$, on the population depicted by $f$. In terms of the dimensionless momentum, the one dimensional operator is:

\begin{equation}
\label{eq:Ldf}
\textsl{L}\left[f(z)\right] = \underbrace{\frac{\partial f }{\partial t} }_{L1}+  \underbrace{\frac{z \sqrt{2m/\beta}}{m} \frac{\partial f }{\partial x}}_{L2}+ \underbrace{\mathcal{F} \sqrt{\frac{\beta}{2m}} \frac{\partial f }{\partial z }}_{L3}
\end{equation}
\noindent
Where, since we will deal separately with each piece of such an operator, we label each term as $L1$, $L2$ and $L3$, respectively. Concerning the right hand side of Eq.(\ref{eq:BEq-gral}), the collision operator, here is taken in accordance with the BGK operator, Eq.(\ref{eq:BGK}). As above mentioned,  the explicit form we use here to handle such an operator differs from the usual way, and it shall be explained through the next section. As previously stated, within the usual BGK  term,  the perturbed distribution is represented by $f$, and the $f_0$ one stands for the equilibrium.  Since we deal with it in further sections, here we write its explicit form:
\begin{equation}
\label{eq:BGK}
 \textsl{I}_C\left[ z \right] \equiv - \frac{f_{0} - f(z)}{\tau_c} \equiv - \frac{\Delta f(z)}{\tau_c}
\end{equation}

\section{The Central Limit Theorem}

The $\Delta f(z)$ factor within the BGK operator, Eq.(\ref{eq:BGK}), from the Physics standpoint,  can be regarded as the departure of an ensemble of charges from the equilibrium  with respect to the relaxation time. Such an equilibrium, $f_0$,  is described by means of an isotropic Gaussian, \cite{Reif}, i.e. the  Maxwell-Boltzmann distribution, or also an equilibrium Fermi-Dirac distribution, \cite{Lindsay}. The population out of equilibrium, $f$, usually is depicted either by means of a  small perturbation factor, $\phi$, as $f = f_0(1+\phi)$ \cite{Lindsay, Reif, Zhdanov}.  Furthermore, one  also can use the Kramers approach \cite{Alekseev}, or other distributions, as the kappa distribution, which corresponds to a solution of the Fokker-Planck equation regarding collisional processes and collective effects \cite{Hasegawa}. Throughout this work, the population out of the equilibrium is depicted by means of a right-skewed power-law distribution. The main reason of that choice, comes up from the fact that such a distribution is an exact solution of the Beq. \cite{Kats1}. On the other hand, from the mathematical standpoint, the equilibrium distribution is a Gaussian, and therefore  the Central Limit Theorem \cite{CLT, CLT2} can be used: For a given cumulative distribution, $f$, the departure from the Gaussian distribution, $\Delta f$,  can be  written in terms of the statistic moments up to the $4th$ order  as follows, \cite{Tchevyshef}:  

\begin{equation}
\label{eq:TCL}
\Delta f(z)=\frac{1}{\sqrt{2\pi}}\  Exp\left[-z^2/2\right] \left[ \frac{Q_1(z)}{\sqrt{N_f}} + \frac{Q_2(z)}{N_f}+ \cdots \right]
\end{equation}
\noindent 
Here $N_f$ stands for the number of degrees of freedom. The $Q$-terms include the contributions of the $3rd$ Moment, $M_3(z)$ (the  skewness of the distribution), and the $4th$ moment, $M_4(z)$ (the kurtosis).  In Eq(\ref{eq:TCL}) the $Q_1$ and $Q_2$ terms are,

\begin{equation}
\label{eq:Q1}
Q_1(z) = \frac{M_3(z)}{6} \left[  z^2 - 1\right] \\
\end{equation}

\begin{eqnarray}
\label{eq:Q2}
Q_2(z) = \frac{z^5}{72} \left( M_3(z) \right)^2 + \frac{z^3}{24} \left[ M_4(z) - \frac{10}{3} \left( M_3(z) \right)^2 \right] + \\ 
+ \frac{z}{8} \left[ \frac{5}{3} \left( M_3(z) \right)^2 - M_4(z) \right] \ \ \ \ \ \ \ \ \ \  \nonumber
\end{eqnarray}
For practical applications,  the deviations from the asymptotic behavior can be seen even  when one sums up a large but \emph{finite} number of  variables. The study of such an issue,   with a sufficiently regular distribution ensuring the existence of higher moments, can be found in \cite{CLT}. In our case, the moments of the power-law distribution, follow  Eq.(\ref{eq:PL-Mom}). At this stage, having all the pieces of Eq.(\ref{eq:BEq-gral}), we apply it to analyse the evolution of a plasma taking into account such a ponderomotive force. Within the next section, we solve such a BEq.

\section{The Beq using the pulsed oscillating field and the ponderomotive force}
\label{Beq1}
As explained, here we introduce the considered forces $\mathcal{F}(x,t)$  into the Beq during the pulse $g_0(t)$, according to Eq.(\ref{eq:Forces1}). First, we handle the l.h.s. of Eq.(\ref{eq:BEq-gral}), the $\textsl{L}\left[f(z)\right]$ operator on $f(z)$. According to Eq.(\ref{eq:Ldf}), we have three pieces to calculate. The $L_1$ piece  vanishes since the power-law distribution is not explicitly time dependent. Concerning the $L2$ term  of Eq(\ref{eq:Ldf}), we develop the derivative keeping in mind the temperature is a local value, within the dimensionless $z=p/p_0$ momentum, Eq.(\ref{eq:pz}). Therefore, we write,

\begin{equation}
\frac{\partial f }{\partial x}=\underbrace{\frac{\partial f }{\partial T}}_{(a)}\underbrace{\frac{dT}{dx}}_{(b)}
\label{DTermB}
\end{equation}
\noindent
Using the power-law, $f(z)$, Eq.(\ref{eq:PLDist}), and the $z$ definition, the term labelled as $(a)$ reads,

\begin{equation}
\frac{\partial f }{\partial T}  = \frac{- \mu\ f }{ T }
\label{Dfrz}
\end{equation} 
\noindent

Concerning the above $(b)$ term, the temperature gradient, $dT/dx$, it is related with the electric field, $\mathcal{E}_0$, and it represents a thermodynamic force per unit charge along the time pulse $\tau_0$. Since we can write $\mathcal{E}_0\sim (kB/q)dT/dx$,  \cite{Lindsay, Ziman}, then, the  $(b)$ term $\sim\ q \mathcal{E}_0/k_B \sim p/\tau_0$. By writing it as a function of $z$:

\begin{equation}
\frac{dT}{dx} \sim\ F_e/k_B \sim \frac{1}{k_B} \frac{\sqrt{2m/\beta}\ \ z }{\tau_0} 
\label{gradT}
\end{equation} 
Hence, merging  Equations, (\ref{Dfrz}) and (\ref{gradT}) within Eq.(\ref{DTermB}), the $L2$ piece of the operator,  Eq.(\ref{eq:Ldf}), yields:

\begin{equation}
\frac{z \sqrt{2m/\beta}}{m} \frac{\partial f }{\partial x}=\ z^2\ \frac{(-2\mu\ f)}{\tau_0} 
\label{TERMB}
\end{equation}  
Concerning the $L3$ term of Eq.(\ref{eq:Ldf}), the  force  $\mathcal{F}(x,t)$, comes from Eq.(\ref{eq:Forces1}). The $L3$ term reads,
\begin{equation}
\mathcal{F}\  \sqrt{\frac{\beta}{2m}} \frac{\partial f }{\partial z } =\ \frac{-2\mu f }{z}\  \mathcal{F}\  \sqrt{\frac{\beta}{2m}}   
\label{TERMC}
\end{equation}
Hence, by merging Eqs.(\ref{TERMB}), (\ref{TERMC}) into Eq.(\ref{eq:Ldf}), we attain the $\textsl{L}\left[f(z)\right]$ operator,

\begin{equation}
\textsl{L}\left[f(z)\right] =\ \ \left(-2\mu f\right)\ z^2 \left[ \sqrt{\frac{\beta}{2m}} \frac{\mathcal{F}}{z^3} + \frac{1}{\tau_0} \right]
\label{eq:Ldf2}
\end{equation}

Following Eq.(\ref{eq:BEq-gral}),  we need to equate  Eq.(\ref{eq:Ldf2}) with the  $\textsl{I}_C\left[ z \right]$ term, Eq.(\ref{eq:BGK}). To find the explicit expression of the latter operator, as explained,  we use the Limit Central Theorem,  Eq.(\ref{eq:TCL}), up to the term $Q_2(z)$, together with Eq.(\ref{eq:PL-Mom}). We obtain:  

\begin{eqnarray}
\label{eq:BGK2}
 \textsl{I}_C\left[ z \right]= \frac{-\Delta f(z)}{\tau_c}= \frac{-1}{\tau_c} \frac{Exp\left[-z^2/2\right]}{\sqrt{2\pi}} \biggl[ \frac{\left( z^2 - 1 \right) }{6} \frac{M_3(z)}{\sqrt{N_f}} + \\  
+ \frac{\left( z^5 - 10 z^3+ 15 z \right)}{72} \frac{\left( M_3(z) \right)^2}{N_f} + \frac{\left(z^3 - 3 z \right)}{24} \frac{ M_4(z)}{N_f} \biggr]  \nonumber
\end{eqnarray}
\noindent 
Where
\[
 M_3(z) =\frac{(2\mu -1) N_f}{2\mu -4} ;\   M_4(z) =\frac{(2\mu -1) N_f}{2\mu -5}
\]

Finally, using the definition of $f(z)$, Eq.(\ref{eq:PLDist}), within Eq.(\ref{eq:Ldf2}) and equating it with  Eq.(\ref{eq:BGK2}), we obtain the relationship between the  characteristic time $\tau_c$ and the time pulse  $\tau_0$:

\begin{eqnarray}
\label{eq:QuotT}
\tau_c =  \left[ \frac{1}{z^3} \frac{\mathcal{F}}{p_0} + \frac{1}{\tau_0} \right]^{-1}
\frac{Exp\left[-z^2/2\right]}{\sqrt{2\pi}} \frac{z^{2(\mu -2)}}{2 \mu} \biggl[ \frac{\left( z^5 - 10 z^3+ 15 z \right)}{72} \frac{2\mu-1}{(2\mu-4)^2}\  + \\
+\ \frac{z^2 - 1 }{6 (2\mu-4) \sqrt{N_f}}\ +\ \frac{z^3 - 3 z}{24 (2 \mu -5) N_f} \biggr]  \nonumber 
\end{eqnarray}
As  we can see, the first term does not depend on $N_f$, and the last two terms are order $O(1/(\sqrt{N_f})$ and  $O(1/(N_f))$ respectively, then for large $N_f$, the leading term  is first one. This weak dependence comes from the cancellation between the term  of the plasma density normalisation of the $f(z)$ distribution function, and the corresponding density factor within the  different moments.   
Here we must point out, in despite that at first glance this expresion  provides a result which  we could consider that is  almost $N_f$ independent for large $N_f$,  in a realistic computation, such a dependence would be still present, as the plasma frequency depends on the actual plasma density. 

At this stage, we will discuss the domain of application of the equation Eq.(\ref{eq:QuotT}) using the statistical moments of the power-law distribution: As established above, such an equation has been obtained from Beq, Eq.(\ref{eq:BEq-gral}), in which we take a BGK collision operator, Eq.(\ref{eq:BGK}) within the context of BG statistics. Such an operator involves the classical CLT through the term $\Delta f(z)$, using a power-law distribution, which allows us to extract the relaxation time in a relatively simple way. In classical CLT, the random variables are required to be independent, and the CLT does not hold if the correlation between long-range random variables is non-negligible \cite{Tsallis2}. Therefore, this fact limits the domain of application of the obtained expression. Relaxation time effects arising from long-range correlations cannot be treated in the plasmas considered here. In reference \cite{Tsallis2}, an extension of classical CLT using  the q-algebra, considering long-range correlations, is considered. Considering future research, perhaps if a similar but non-extensive BGK-type collision operator could be constructed that was coherent in the context of BEq, then perhaps the relaxation time considering the effects of such correlations could be obtained in a similar way.\\

Within next sections  we will test  the above expresion and parameterise it to study the impact of the relevant physical quantities. On the other hand, as expected, from the above equation, the relaxation time of the plasma depends on the explicit form of the interacting force,  $\mathcal{F}$, acting during the pulse. Moreover, there is a dynamic dependence through the dimensionless momentum $z(t)= p(t)/p_0$, therefore, in order to include such a time dependence of $z(t)$, first we need to solve the equation of motion of the charges under the action of the pulse $\mathcal{F}(x,t)$. 

\section{The Equation of motion}

The solution of  the equation of motion of this particular system of a gaussian pulse containing the above considered  forces, to the best of our knowledge, is not present in the literature,  therefore, to obtain $z=z(t)$, we must solve:
 
\begin{eqnarray}
\mathcal{F}(x,t) &=& \frac{dp}{dt}= \left[ F_{se}(x)\ cos\left( \omega t \right) -\ F_{pdm}\right]\  g_0(t); \! \! \! \! \! \\
g_0(t) &\equiv& \sum_{k=0}^N Exp\left[ -\left(\frac{t - 2kt_0}{t_0}\right) ^2\right]  
\label{eq:Forces2}	
\end{eqnarray}
\noindent 
By integrating the Eq. of above we write,

\begin{equation}
p(t) =\  F_{se} I_a(t) - F_{pdm} I_b(t) + Constant
\label{eq:Pdt1}	
\end{equation}
\noindent
where 
\begin{equation}
I_a(t) =\  \sum_{k=0}^N I_{a,k}(t) \equiv \sum_{k=0}^N \int Exp\left[ -\left(\frac{t - 2kt_0}{t_0}\right) ^2\right] cos\left( \omega t \right) dt
\label{eq:Ia1}	
\end{equation}
and 
\begin{equation}
\label{eq:Ib1}
I_b(t) = \int \ g_0(t)\ dt =\  \frac{\sqrt{\pi}}{2}t_0 \sum_{k=0}^N Erf\left[ \frac{t-2kt_0}{t_0}\right]  
\end{equation}
where $Erf[ \ \ ]$ stands for the Error function of the argument, \cite{Abramovitz}.  Solving Eq.(\ref{eq:Pdt1}), we set as the initial condition, at $t=t_0$ that the pulse begins upon a thermalised population  $p_0^2=2m/\beta$. Hence, we  rewrite Eq.(\ref{eq:Pdt1}) as,

\begin{equation}
p(t) =\ p_0 + \ F_{se} \left[ I_a(t)-I_a(t_0) \right] - F_{pdm} \left[ I_b(t)-I_b(t_0) \right]
\label{eq:Pdt2}	
\end{equation}
\noindent
Therefore, we need to compute the $I_{a,k}(t)$ integral within Eq.(\ref{eq:Ia1}). The full details about such a calculation can be found in  \ref{App1}, here we only write the final result: By performing the following changes,

\begin{equation}
 x = \frac{t - 2k t_0}{t_0} ;\ \  y=  \omega t_0;\ \ a_k=2ky;\ \ xy = \omega(t - 2k t_0)
\label{eq:chng}
\end{equation}
after calculations we attain,

\begin{eqnarray}
 I_{a,k}(t) \equiv   t_0\ \left[ Re( \breve{I}_{a,k}(t) ) - i\ Im( \breve{I}_{a,k}(t)) \right] =  \ t_0\ \frac{\sqrt{\pi}}{2} Exp\left[-y^2/4 \right]\times \\ \nonumber
\times\ \left\{ cos( a_k )\ \left[ Erf(x)+ \frac{Exp\left[-x^2 \right]}{2 \pi x} \left[ ( 1-cos(x y) ) \right] \right] -\ i\ sin( a_k ) \frac{Exp\left[-x^2 \right]}{2 \pi x}  sin(xy) \right\} 
 \! \! \! \! \! \! \! \! \! \! 
\label{eq:Iak4}	
\end{eqnarray}

and finally, taking into account that $x, xy$, and $a_k$ are $k$-dependent, and extracting for convenience the $t_0$ time, we write
\begin{equation}
I_a(t) =\  \sum_{k=0}^N I_{a,k}(t) =   t_0\ \sum_{k=0}^N \left[ Re( \breve{I}_{a,k}(t) ) - i\ Im( \breve{I}_{a,k}(t)) \right]
\label{eq:Ia2}
\end{equation}

\subsection{The dimensionless $z(t)$ momentum} 

\begin{figure}
\center
\includegraphics[width=8.5cm]{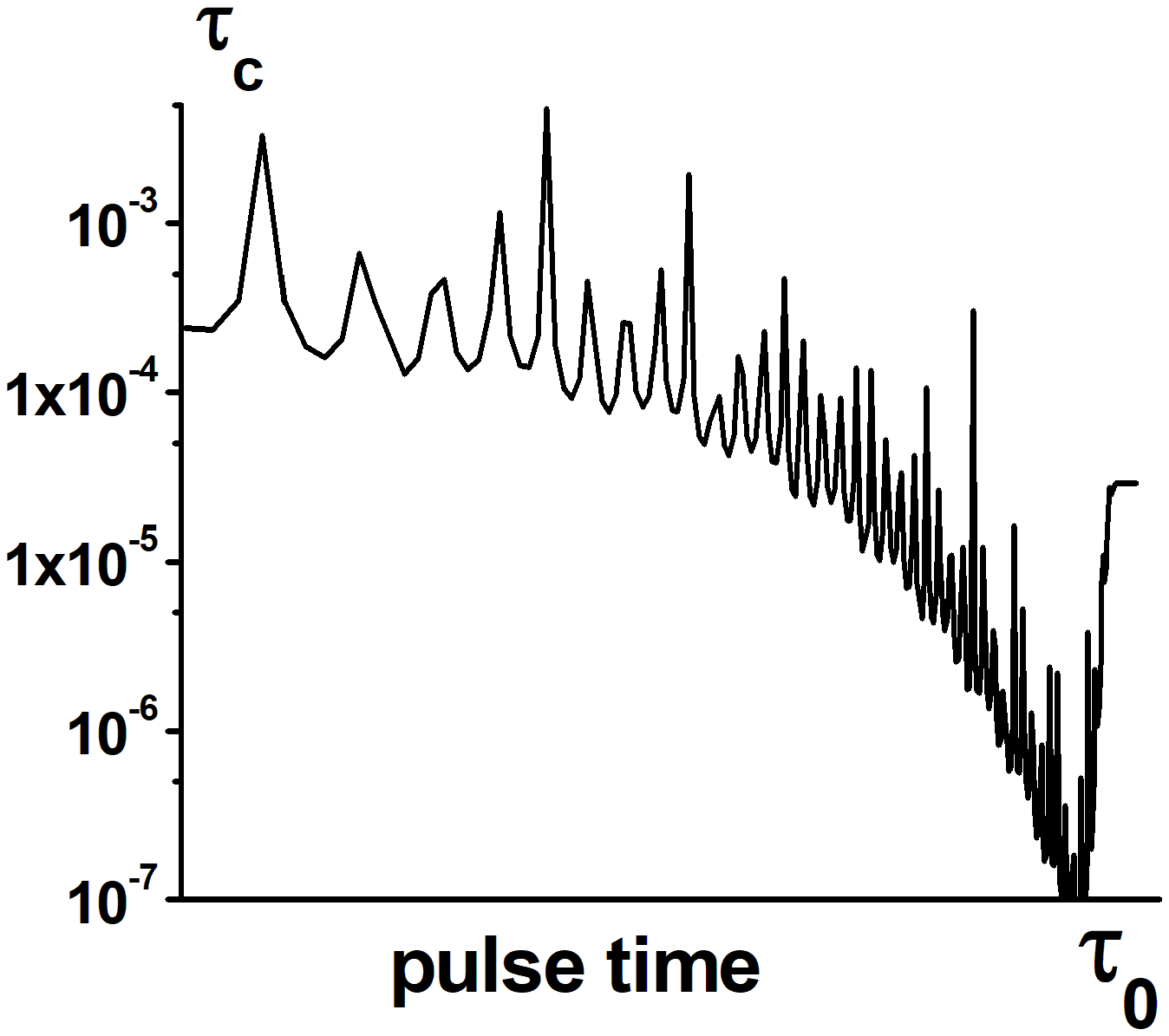} 
\caption{\label{fig3:TZt} Evolution of  the relaxation time,$\tau_c$, along the time pulse, $\tau_0$ provided by the Eq.(\ref{eq:QuotT}) using the Central Limit Theorem and the dimensionless z(t) according to Eq.(\ref{eq:zdt1}). Parameters are (see Section \ref{Sect:Para}): $N_f=10^{10}$; $ \mu=4.6 $; $ N=10 $; $ \alpha= 6 $. }
\end{figure}

Next, here we will write the $z(t)$ momentum in terms of the real and imaginary part.  First we extract the $t_0$ time from $I_a(t)$ and $I_b(t)$, Eqs. (\ref{eq:Ia2})  and (\ref{eq:Ib1}), respectively. We rename the latter integrals as  $I_a(t)\equiv t_0 I_1$ and $I_b(t) \equiv t_0 I_2$. Then, within Eq.(\ref{eq:Pdt2}) we can write both the inhomogeneous and ponderomotive force in terms of the momentums: $F_{se} t_0 \equiv P_{se}$; $F_{pdm} t_0 \equiv P_{pdm}$. Therefore, dividing Eq.(\ref{eq:Pdt2}) by $p_0$, we finally write the equation of motion in terms of the dimensionles momentum as,

\begin{equation}
z(t) =\ 1 + \ Z_{se}  \left[ I_1(t)-I_1(t_0) \right] - Z_{pdm}\left[ I_2(t)-I_2(t_0) \right]
\label{eq:zdt}	
\end{equation}
\noindent
Where, $Z_{se}\equiv P_{se}/p_0$ and $Z_{pdm}\equiv P_{pdm}/p_0$. As the $I_1(t)$ integrals, according to Eq.(\ref{eq:Iak4}) are complex numbers, we can separate the dimensionless momentum into the real and imaginary part.  

\begin{eqnarray}
z(t) =\ 1 + \ Z_{se}  \left[ Re(I_1(t))-Re(I_1(t_0)) \right] - Z_{pdm}\left[ I_2(t)-I_2(t_0) \right] +  \\ 
+\ i\ Z_{se}  \left[ Im(I_1(t))-Im(I_1(t_0)) \right] \ \ \ \ \ \ \ \ \ \ \ \ \ \ \ \ \ \  \ \nonumber
\label{eq:zdt1}	
\end{eqnarray}
\noindent
In Figure \ref{fig3:TZt} , we can see the solution of Eq.(\ref{eq:QuotT}) in which, the dimensionless z(t) is  the real part  of Eq.( \ref{eq:zdt1} )

\subsection{Parameterisation of the forces acting on the relaxation time}
\label{Sect:Para}
In this section, we present the parameters that we will use to perform the numerical estimates. As mentioned above, two main forces are involved: the local electromagnetic oscillating force $F_{se}(x)\ cos(w t)$ of an inhomogeneous electric field and the Miller ponderomotive force, which we express in terms of the plasma frequency, with the dielectric constant being unity, as $F_{Pdm}= -(\omega_p^2/2\omega^2) \nabla(\textbf{$\mathcal{E}$}^2)$. Regarding the frequency, $\omega$, to perform these estimates, we take a parameter, $\alpha$, with respect to the plasma frequency, $\omega_p$, which is taken as a reference. The parameter $\alpha$ is then read as $\alpha\ =\omega/ \omega_p$, and here, for simplicity, $\omega_p$ is set to one. We then parameterize the $\mathcal{F} / p_0$ term within equation (\ref{eq:QuotT}). According to equation (\ref{eq:Forces1}), and following the discussion in the previous section on dimensionless Z, we write:
\begin{eqnarray}
\frac{\mathcal{F}(x,t)}{p_0} =  \left[ Z_{se}\ cos\left( \omega t \right) - Z_{pdm}\right]\  g_0(t); \! \! \! \! \! 
\label{eq:param1}	
\end{eqnarray}
\noindent 
Here, $Z_{pdm}\sim C_{pdm} \times(\omega_p^2/\omega^2)=C_{pdm}/\alpha^2$, where $C$ is a constant depending on the gradient of the average  square electric field. Furthermore, the relative strength of the dimensionless electric field $Z_{se}$ with respect to $Z_{pdm}$ will be considered in a later study. Other parameters are: the power law coefficient, $\mu$; $N_f$; and $N$ is the number of pulses along the time pulse $\tau_0$.
\begin{figure}
\center
\includegraphics[width=8.5cm]{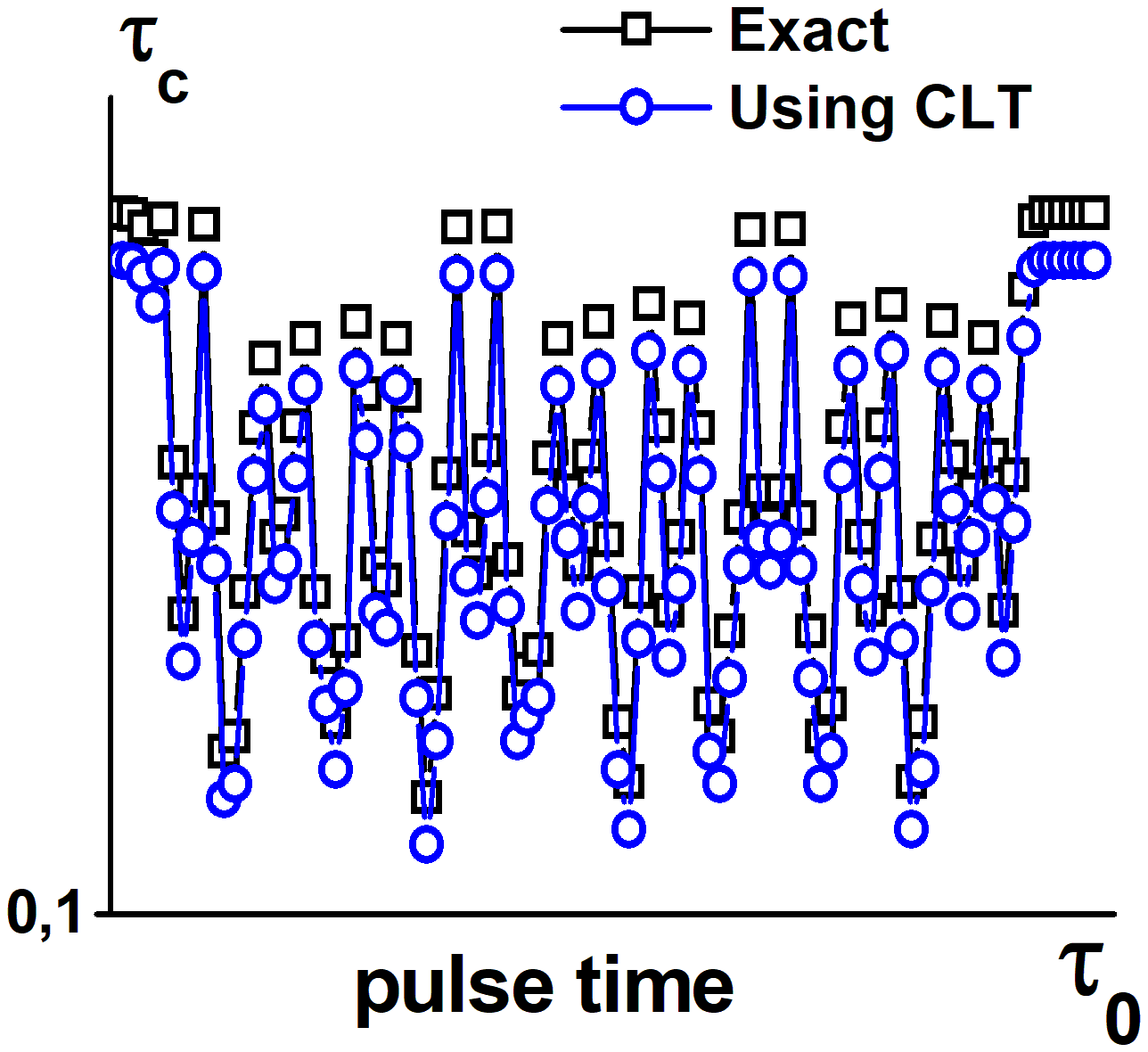} 
\caption{
\label{fig4:Tau-Comp}  Comparison of the solution of the relaxation times  obtained by the Eq.(\ref{eq:QuotT}) using the Central Limit Theorem (circles) and the corresponding exact BGK  solution provided by Eq.(\ref{eq:ExQuotT}), using the same parameters, (squares). Parameters are (see Section\ref{Sect:Para}): $N_f=10^{10}$; $ \mu=3.1 $; $ N=20 $; $ \alpha= 5 $. }
\end{figure}  

\section{Test of the obtained equation  governing the relaxation time}
To test the developed equation (Eq.(\ref{eq:QuotT}) using the statistical moments of the power-law distribution $f(z)$, we compare it with the "exact" BGK collision operator, obtained directly as the difference $f_0-f(z)$, Eq.(\ref{eq:BGK}), for that distribution. We then calculate the exact expression by taking Eq.(\ref{eq:BEq-gral}) by inserting the operator $\textsl{L}\left[f(z)\right]$, Eq.(\ref{eq:Ldf2}), inside. We denote the relaxation time obtained in this way as $(\tau_c)_{exact }$. After a similar calculation, we attain:

\begin{eqnarray}
\label{eq:ExQuotT}
(\tau_c)_{exact} =  \left[ \frac{1}{z^3} \frac{\mathcal{F}}{p_0} + \frac{1}{\tau_0} \right]^{-1}\ \frac{1}{2 \mu z^2}\biggl[ 1 - \frac{ Exp\left[ -z^2/2 \right]\ z^{2\mu}}{\sqrt{2\pi}\ (2\mu-1)} \biggr]
\end{eqnarray}
We then insert the same parameters to evaluate both equations, (\ref{eq:QuotT}) and (\ref{eq:ExQuotT}). In Figure \ref{fig4:Tau-Comp} we can see the result. In order to compare we  insert the same parameterised $z$ values. Other  parameters can be read within the caption. In this case the result yield a relative diference between both results which is $\sim 6\%$. The most relevant value is the $ \mu $ coefficient of the Power law. Nevertheless, by inserting other several educated set of parameters, greater differences, around $\sim 30 \%$, can be found.  
\begin{figure}
\centering
\includegraphics[width=8.5cm]{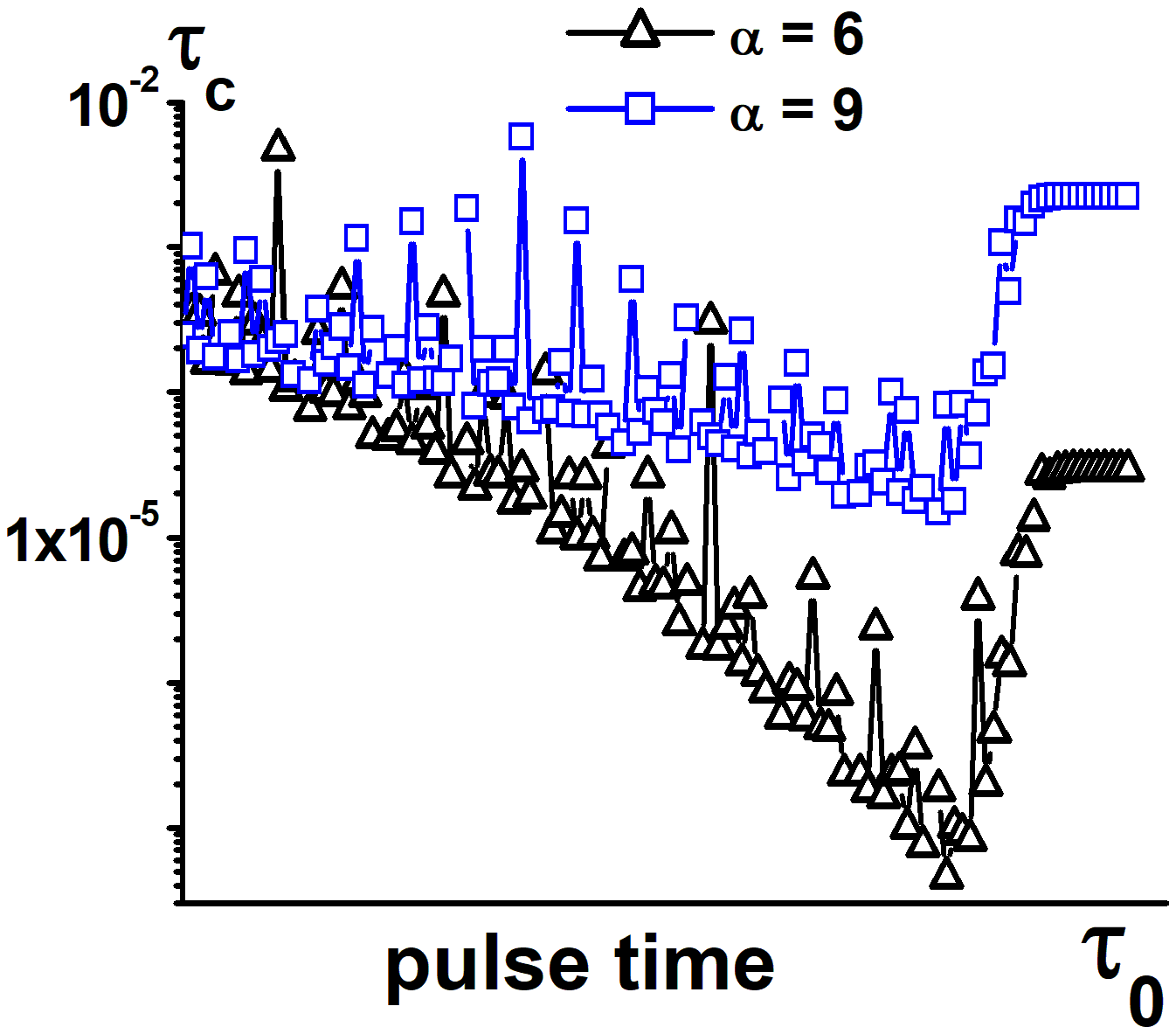}  
\caption{\label{fig5:Tau-w} Estimates of the relaxation time as a function of a parameterised frequency $\omega= \alpha\ \omega_p$. The solution of the relaxation times is provided by the Eq.(\ref{eq:QuotT}) together with the dimensionless z(t) according to Eq.(\ref{eq:zdt1}). Other parameters are: $N_f=10^{10}$; $ \mu=4.6 $; $ N=10 $.}
\end{figure} 

\section{Estimates of the relaxation times}

Finally, in this section, we provide some numerical estimates of the impact of the relevant quantities on the relaxation times, $\tau_c$, according to Eq. (\ref{eq:QuotT}) together with the dimensionless $z(t)$, according to Eq. (\ref{eq:zdt1}), by taking the real part of the dimensionless momentum. The parameters we use here are described in previous section \ref{Sect:Para}. As mentioned, these numerical estimates aim only to examine the impact of relevant physical parameters on the relaxation time given by equation (\ref{eq:QuotT}).  A simulation describing the complete physics of a space plasma is beyond the scope of this work: this would imply defining a complete set of parameters of that specific plasma, the actual boundary conditions, and, in addition, the actual conditions of the external pulsed electromagnetic field that perturbs it.

The first estimate concerns  the response of $\tau_c$ with respect to the frequency parameter, $\alpha$. The result  can be observed in Figure \ref{fig5:Tau-w}, in which we find the plot corresponding to the behaviour of the evolution of  the relaxation time,$\tau_c$, along the time pulse, $\tau_0$, as a function of a parameterised frequency $\omega= \alpha\ \omega_p$. The solution of the relaxation time is provided by the Eq.(\ref{eq:QuotT}) using the Central Limit Theorem by inserting  the dimensionless z(t) according to Eq.(\ref{eq:zdt1}). The common values of the other relevant parameters can be seen within the caption. From Figure we realise that from relative low frequencies with respect to the plasma $\omega_p$ there is a fast relaxation time decay  along with the time pulse. It decreases orders of magnitude along the time pulse  with respect to the values at the beginning  and later, at the end of the pulse, the excitation decreases going towards the initial values. The rapid oscillation of that time during the intermediate gaussian pulses suggest a complex  kinetic mechanism of interaction which could be a fast momentum interchange  between the plasma population due to the interplay between the frequency of the oscillating field and  the ponderomotive force during two consecutive gaussian pulses.  On the  other hand, at relative higher frequencies, we notice that the relaxation time decreases slowly along the pulse but the rapid oscillation  during the intermediate gaussian pulses are still present. Looking for a posible explanation on that behaviour, by reading Eq.(\ref{eq:QuotT}), and the force interaction within the equation of motion,  we should then  study the ponderomotive term and its relative strengh with respect to the $Z_{es}$:  Acording to discussion within section \ref{Sect:Para}, such a term reads $Z_{pdm}\sim C_{pdm}/\alpha^2$. in terms of the physical parameters, the ponderomotive force,is frecuency dependent, through $\alpha$, and also it depends on the $C_{pdm}$ parameter,  the gradient of the electric field.  Both factors are present during the consecutive gaussian pulses untill the pulse ends. Concernig the comparison between relative high and low  frequency, the lower frequencies means actually we are increasing the strength of $Z_{pdm} \sim 1/\alpha^2$  with respect to the $Z_{es}$. Conversely, higher frequencies means to make $Z_{pdm}$ weaker  and $Z_{es}$ dominates. We can test the suggested balance is mainly responsible of the relaxation time behaviour by means the study of the the $C_{pdm}$ parameter, the averaged squared electric field gradient. 
\begin{figure}
\center
\includegraphics[width=8.5cm]{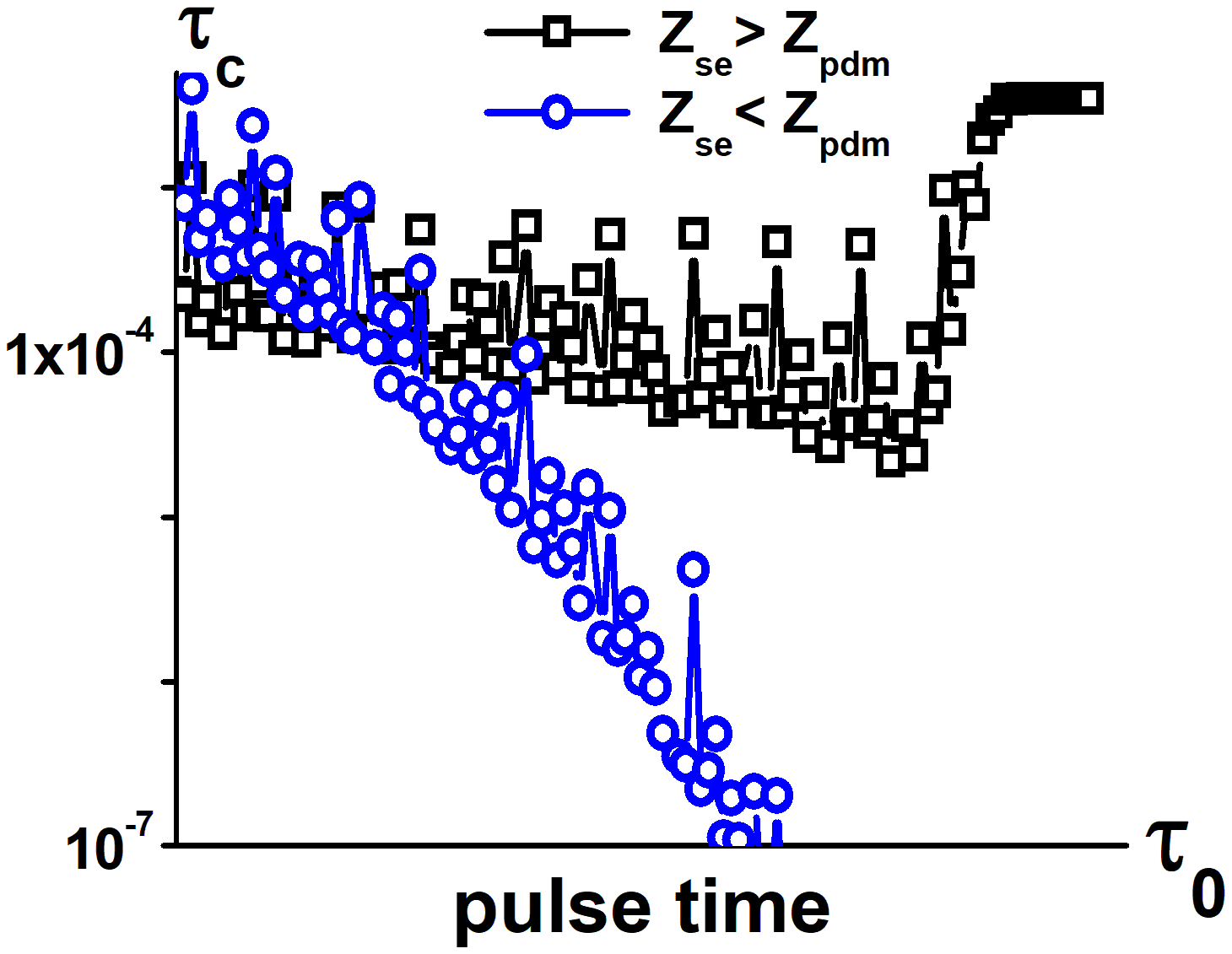} 
\center
\caption{\label{fig6:TZstr} Relaxation time along a time pulse by the interchange of the value of the strength of the dimensionless $Z$ parameter of the electromagnetic force $Z_{es}$ and the ponderomotive factor $Z_{pdm}$, keeping the rest of parameters. $\alpha=10$ ;$N_f=10^{10}$; $ \mu=4.6 $; $ N=10 $. Squares: The $Z_{es}> Z_{pdm}$. Circle: the numerical values are interchanged, $Z_{pdm} > Z_{es}$. } 
\end{figure}  
We then display  the relative strength of the dimensionless parameter of the electromagnetic force $Z_{es}$ with respect to the factor $C_{pdm}$. Figure \ref{fig6:TZstr} shows the evolution of the Relaxation time along a time pulse by the interchange of the value of the strength of the dimensionless $Z$ parameter of the electromagnetic force, $Z_{es}$, and the ponderomotive factor, $C_{pdm}$, by keeping the same frequency, and as well the rest of parameters. the graph with squares refers to the case in which  $Z_{es} > C_{pdm}$. In the plot with circles the numerical values are interchanged from the previous case, giving  $C_{pdm} > Z_{es}$. We can observe the fast decay along the time for  the $Z_{pdm}$  dominance. We can then conclude that both, the frequency of the field acting ond the plasma and  also the gradient of the squared electric field governs the balance between the, $Z_{es}$, and the ponderomotive factor, $Z_{pdm}$. A full simulation of the detailed kinetic mechanism  of the momentum interchange by concatenating successive gaussian pulses looking on such $Z_{pdm}$ factors could be of interest from both  the theoretical and experimental point of view.  

\section{Conclusions.} 
In this work, we studied the relaxation time of a plasma which is perturbed by means of  a time dependent square pulsed force. Such a pulse is built by means of a Gaussian superposition. During such a pulse two forces are considered: An inhomogeneous  oscillating electric force and the corresponding ponderomotive force. The study of the evolution of the ensemble is carried out by the Boltzmann Equation in which, concerning the collision term,  we present a new procedure: Instead the usual techniques, to  take into account  the difference between the equilibrium and non-equilibrium distributions, $\Delta f$, here the Central Limit Theorem is used, and such a $\Delta f$ term is written as a function of the statistical moments. We afford an explicit expression of such a term using a power-law  distribution function for the perturbed population, which is an exact solution of the Beq.. This procedure appears to be a reliable alternative for use in the absence of a specific collision term within the Beq . Furthermore, as far as the forces involved are concerned, the same procedure presented here can be extended to many scenarios. We use this technique with the Gaussian pulse of an inhomogeneous oscillating electric force and the ponderomotive force. Since the solution of the equation of motion of this particular system is, to our knowledge, not present in the literature, we provide the solution of the corresponding equation of motion in terms of the dimensionless momentum.  Moreover, we   provide the explicit expression of the evolution equation of  the relaxation time with respect to the time pulse as a function of the dynamic conditions by inserting the ad hoc solved dimensionless momentum equation. Next, we discuss the scope and limitations of the obtained expression for the relaxation time.

In order to test the developed equation Eq.(\ref{eq:QuotT}) using the statistical moments of the Power-law, $f(z)$ distribution, we compare it with the corresponding expresion using the exact collission BGK operator directly  obtained as the diference $f_0-f(z)$ , Eq.(\ref{eq:BGK}), for such a distribution. To achieve this goal we solve  the exact expresion, and we compare the relaxation time evolution from  both obtained expressions, by using the same parameterisation. 

 In addition, we provide numerical estimates of such relaxation times using these solutions. These estimates, involves the relaxation times, $\tau_c$, according to Equation (\ref{eq:QuotT}) together with the  dimensionless $z(t)$, according to Eq.(\ref{eq:zdt1}). We devote the estimates to a generically parameterised  plasma,  looking at the physical parameters as the frecuency, $\alpha$, and as well the the gradient of the squared electric field of the pondermotive force, ($C_{pdm}$) . We find that the evolution of the relaxation time comes from the balance between the electromagnetic term $Z_{es}$ and the ponderomotive factor $Z_{pdm}$, which is governed by  $\alpha$, and $C_{pdm}$. To define a complete set of  parameters for an actual  plasma, with appropriate boundary conditions, and the actual conditions for the  external  pulsed electromagnetic field perturbing the plasma would be the next step to apply this effort to  space plasmas.


\begin{acknowledgments}
We acknowledge Prof. M.A. Sanchis-Lozano the careful reading of the manuscript and their valuable comments.
\end{acknowledgments}

\appendix

\section{Evaluation of the $I_{a,k}(t)$ integral}
\label{App1}

Here, we evaluate

\begin{equation}
 I_{a,k}(t) \equiv \int Exp\left[ -\left(\frac{t - 2kt_0}{t_0}\right) ^2\right] cos\left( \omega t \right) dt
\label{eq:Iak1}	
\end{equation}

To make easier the calculations, we make the following changes:
\begin{equation}
 x = \frac{t - 2k t_0}{t_0} ;\ \  y=  \omega t_0;\ \ a_k=2ky;\ \ xy = \omega(t - 2k t_0)
\label{eq:app-chng}	
\end{equation}
\noindent
Those changes transforms Eq.(\ref{eq:Iak1}) into,
\begin{equation}
 I_{a,k}(t) =\ t_0\ \left\{ cos( a_k ) \int e^{-x^2} cos\left( x y \right) dx - sin( a_k ) \int e^{-x^2} sin\left( x y \right) dx \right\}
\label{eq:Iak2}	
\end{equation}
Using the  tables  from \cite{NG}, the integrals of above can be written in terms of the Error  function as,

\begin{eqnarray}
\int e^{-x^2} cos\left[x y \right] dx= \frac{\sqrt{\pi}}{4} \int e^{-\frac{y^2}{4}} \left\{ Erf\left[x + \frac{iy}{2}\right] + Erf\left[x-\frac{iy}{2}\right]\right\} \\ 
\int e^{-x^2} sin\left[x y \right] dx=\ i\ \frac{\sqrt{\pi}}{4} \int e^{-\frac{y^2}{4}} \left\{ Erf\left[x + \frac{iy}{2}\right] - Erf\left[x-\frac{iy}{2}\right]\right\}  
\label{eq:IErf}	
\end{eqnarray}
for simplicity we again rename: $z = x + i y/2$ and  $z^* = x - i y/2$; where the simbol $*$ stands for the conjugate complex. Also we write $Z_1= Erf[z]$; and, as $Erf[z^*]=(Erf[z])^*$,  \cite{Abramovitz}, we can also write $Z_1^*= (Erf[z])^*$. By inserting the integrals of Eq.(\ref{eq:IErf}) and the latter changes within Eq.(\ref{eq:Iak2}), we attain,
\begin{eqnarray}
 I_{a,k}(t) =\ t_0\ \frac{\sqrt{\pi}}{4} e^{-\frac{y^2}{4}}  \left\{ cos( a_k ) [Z_1 + Z_1^*] -\ i\ sin( a_k ) [Z1 - Z1^*] \right\} =\\
=\ t_0\ \frac{\sqrt{\pi}}{4} e^{-\frac{y^2}{4}}  \left\{ cos( a_k )\ 2\ Re(Z_1) -\ i\ sin( a_k )\ 2\ Im(Z_1) \right\}
\label{eq:Iak3}	
\end{eqnarray}
Where $Re(Z1)$ and $Im(Z1)$ stands for the real and imaginary part of Z1. In order to separate the contributions from the real and imaginary part of the Eq. of above, according to \cite{Abramovitz}, we will expand the Error function as follows,
\begin{equation}
Z_1 = Erf\left[x + \frac{iy}{2}\right] \approx Erf(x)+ \frac{e^{-x^2}}{2 \pi x} \left[ \left( 1-cos(x y) \right) +\ i\ sin(x y)  \right] + \cdots
\label{eq:erf-exp}
\end{equation}
Hence, finally using the above expansion within Eq.(\ref{eq:Iak3}), and by renaming for convenience $I_{a,k}(t)\equiv t_0\ \breve{I}_{a,k}(t)$, the $I_{a,k}(t)$ integral reads

\begin{eqnarray}
 I_{a,k}(t) \equiv   t_0\ \left[ Re( \breve{I}_{a,k}(t) ) - i\ Im( \breve{I}_{a,k}(t)) \right] =  \ t_0\ \frac{\sqrt{\pi}}{2} e^{-\frac{y^2}{4}} \times \\ \nonumber
\times\ \left\{ cos( a_k )\ \left[ Erf(x)+ \frac{e^{-x^2}}{2 \pi x} \left[ ( 1-cos(x y) ) \right] \right] -\ i\ sin( a_k ) \frac{e^{-x^2}}{2 \pi x}  sin(xy) \right\} 
 \! \! \! \! \! \! \! \! \! \! 
\label{eq:app-Iak4}	
\end{eqnarray}

\end{document}